\documentclass[aps,twocolumn,english,reprint, longbibliography, superscriptaddress, breaklinks=true, showkeys, showpacs=false, nofootinbib]{revtex4}
\usepackage[T1]{fontenc}
\usepackage[latin9]{inputenc}
\usepackage{color}
\usepackage{babel}
\usepackage{amsmath}
\usepackage{amssymb}
\usepackage{subfigure}
\usepackage{graphicx}
\usepackage{physics}
\usepackage{comment}
\usepackage[unicode=true,pdfusetitle,bookmarks=true,bookmarksnumbered=false,bookmarksopen=false,breaklinks=true,pdfborder={0 0 0},backref=false,colorlinks=true]{hyperref} 
\makeatletter
\@ifundefined{textcolor}{}
{%
 \definecolor{BLACK}{gray}{0}
 \definecolor{WHITE}{gray}{1}
 \definecolor{RED}{rgb}{1,0,0}
 \definecolor{GREEN}{rgb}{0,1,0}
 \definecolor{BLUE}{rgb}{0,0,1}
 \definecolor{CYAN}{cmyk}{1,0,0,0}
 \definecolor{MAGENTA}{cmyk}{0,1,0,0}
 \definecolor{YELLOW}{cmyk}{0,0,1,0}
}
\pdfoutput=1
\hypersetup{colorlinks=true,citecolor=blue,linkcolor=cyan,urlcolor=blue,filecolor= green, breaklinks=true}
\usepackage{url}
\usepackage{breakurl}
\makeatother

\newcommand{\mf}{\mathfrak}

\begin{document}

\title{Entanglement swapping for partially entangled qudits \\ and the role of quantum complementarity}

\author{Diego S. Starke\href{https://orcid.org/0000-0002-6074-4488}{\includegraphics[scale=0.05]{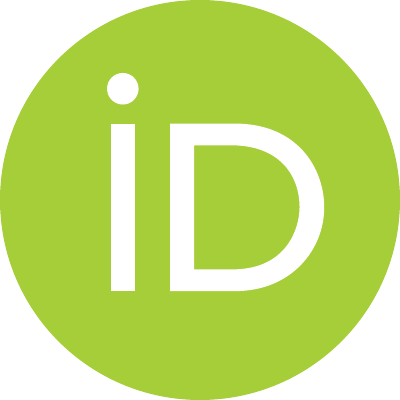}}}
\email{starkediego@gmail.com}
\address{Physics Department, 
Federal University of Santa Maria, 97105-900,
Santa Maria, RS, Brazil}

\author{Marcos L. W. Basso\href{https://orcid.org/0000-0001-5456-7772}{\includegraphics[scale=0.05]{orcidid.pdf}}}
\email{marcoslwbasso@hotmail.com}
\address{Center for Natural and Human Sciences, Federal University of ABC, States Avenue 5001, Santo Andr\'e, S\~ao Paulo, 09210-580, Brazil}

\author{Lucas C. C\'eleri\href{https://orcid.org/0000-0001-5120-8176}{\includegraphics[scale=0.05]{orcidid.pdf}}}
\email{lucas@qpequi.com}
\address{QPequi Group, Institute of Physics, Federal University of Goi\'as, Goi\^ania, Goi\'as, Brazil}

\author{Jonas Maziero\href{https://orcid.org/0000-0002-2872-986X}{\includegraphics[scale=0.05]{orcidid.pdf}}}
\email{jonas.maziero@ufsm.br}
\address{Physics Department, 
Federal University of Santa Maria, 97105-900,
Santa Maria, RS, Brazil}

\selectlanguage{english}

\begin{abstract}
We extend the entanglement swapping protocol (ESP) to partially entangled qudit states and analyze the process within the framework of complete complementarity relations (CCRs). Building on previous results for qubits, we show that the average distributed entanglement between two parties via ESP is bounded above by the initial entanglement of one of the input pairs, and also by the product of the initial entanglements. Notably, we find that using initial states with vanishing local quantum coherence is sufficient to capture the essential features of the protocol, simplifying the analysis. By exploring the cases of qubits and qutrits, we observe that the upper bound on the average distributed entanglement---expressed in terms of the product of the initial entanglements---can be improved, and we conjecture what this tighter bound might be. Finally, we discuss the role of quantum complementarity in the ESP and show how local predictability constrains the entanglement that can be operationally distributed via ESP.
\end{abstract}


\date{\today}

\maketitle

\section{Introduction}

Quantum mechanics, whose core framework emerged around 1925 through the pioneering work of Heisenberg, Schr\"odinger, Dirac, Pauli, Jordan, and Born~\cite{Heisenberg1925,Born1925,Dirac1925, Pauli1925,BornH1926,Born1926,Schrodinger1926,Born1927}, celebrates its centenary this year. This revolutionary theory brought a profound shift in our understanding of atomic and subatomic phenomena. Its development was preceded by foundational breakthroughs, including Planck's quantization of energy~\cite{Planck1900}, Einstein's explanation of the photoelectric effect~\cite{Einstein1905}, Bohr's atomic model~\cite{Bohr1913}, de Broglie's hypothesis of wave-particle duality~\cite{Broglie1924} and Bohr's principle of complementarity~\cite{Bohr1928}. After its formulation, the theory confronted fundamental conceptual challenges such as the EPR paradox~\cite{Einstein1935}, Schr\"odinger's cat thought experiment~\cite{Schrodinger1935}, and ultimately Bell's theorem~\cite{Bell1964}, which formally established a way to experimentally probe the existence of non-local correlations in entangled systems. Building upon this conceptual foundation, quantum theory has enabled the development of quantum computing~\cite{Feynman1982,Lloyd1996}, where superposition and entanglement serve as essential resources, and it has opened pathways for advanced quantum communication protocols~\cite{Simon2017, Ramya2025}, including quantum teleportation~\cite{Bennett1993, Hu2023}. Among these, the entanglement swapping protocol~\cite{Zukowski1993} stands out as a key mechanism for distributing entanglement between distant nodes, thus supporting the quantum networks~\cite{Wei2022,Drmota2023} and the quantum internet~\cite{Kimble2008,Kumar2025,Rohde2025}.

\begin{figure}
    \centering    \includegraphics[width=1\linewidth]{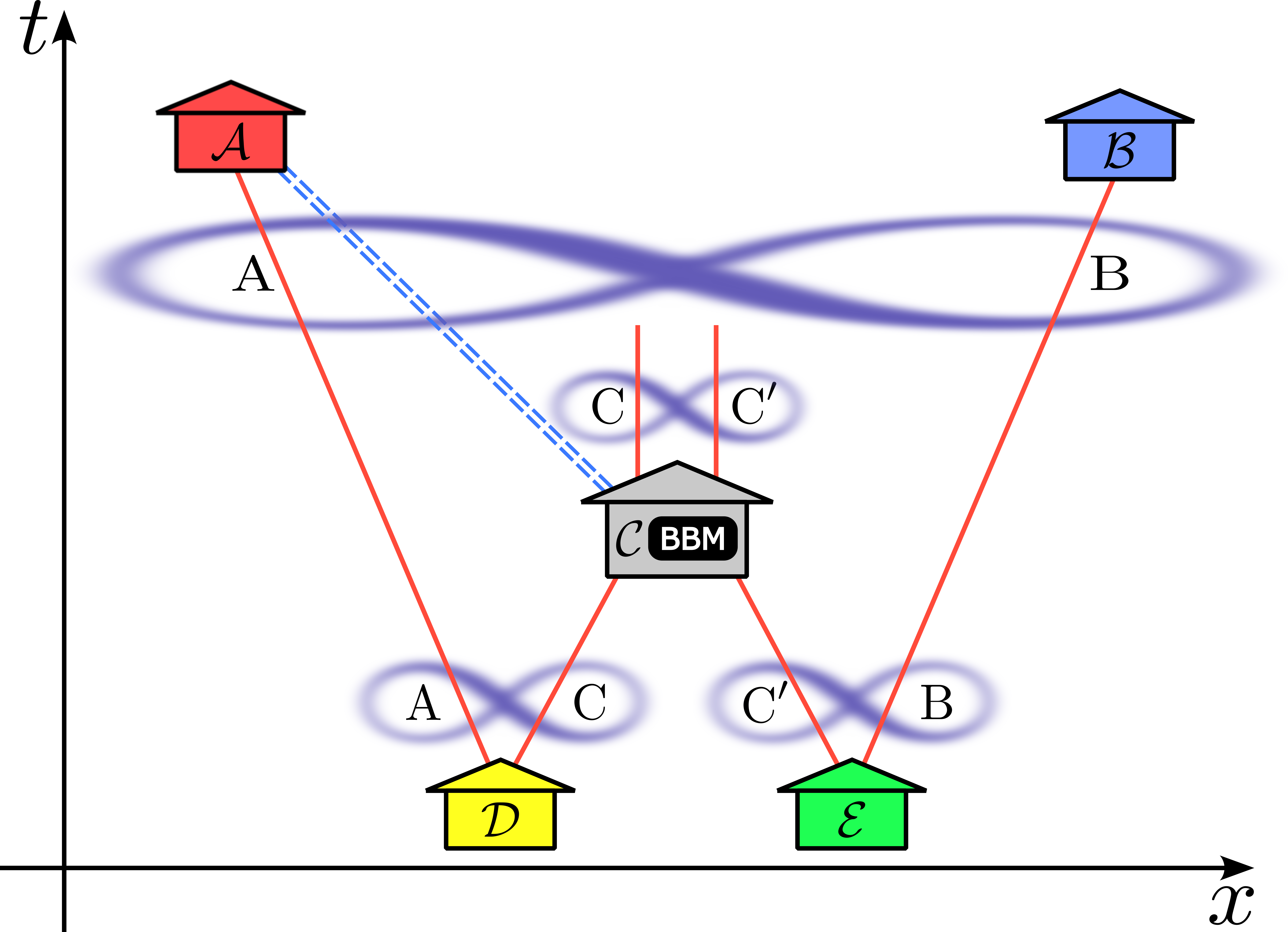}
    \caption{Entanglement swapping protocol in space-time axis can be performed using three laboratories: Alice ($\mathcal{A}$), Bob ($\mathcal{B}$), and Charlie ($\mathcal{C}$). To emphasize the distribution of entanglement over long distances, we can consider two other laboratories: Darwin ($\mathcal{D}$) and Erin ($\mathcal{E}$).  Darwin  sends a pair of entangled quantons ($A$ and $C$) to Alice and Charlie
    , respectively. Erin, in turn, sends the pair of entangled quantons (C$^\prime$ and $B$) to Charlie and Bob
    , respectively. Now, Charlie has qubits from two pairs of entangled quantons: one pair shared with Alice and the other with Bob. Subsequently, Charlie performs a Bell-basis measurement (BBM) on the two qubits in his possession. As a result, Alice and Bob---who initially shared no entanglement---end up with a maximally entangled pair of quantons. The result of the Charlie measurement can be shared by the classical communication channel, represented by the double dashed lines between Charlie and Alice, to allow Alice and Bob to prepare an specific entangled state.
    }
    \label{fig:ESwap}
\end{figure}

Renowned for entangling two quantum systems (or quantons) that have never interacted directly, the entanglement swapping protocol (ESP) was first reported in the seminal work of \.{Z}ukowski \textit{et al.}~\cite{Zukowski1993} and experimentally verified in Ref.~\cite{Pan1998}. A myriad of works on the subject have since been produced, as exemplified in Refs.~\cite{Bose1998,Bose1999,Jennewein2001,Sciarrino2002,Riedmatten2005,Schmid2009,Sangouard2011,Branciard2012,Jin2015,Khalique2015,Tsujimoto2018,Basso2022a,Maziero2023,Davis2025}. ESP can be performed using three laboratories: Alice ($\mathcal{A}$), Bob ($\mathcal{B}$), and Charlie ($\mathcal{C}$). To emphasize the distribution of entanglement over long distances (see Fig.~\ref{fig:ESwap}), Darwin ($\mathcal{D}$) sends a pair of entangled quantons (A and C) to Alice and Charlie, respectively. Erin ($\mathcal{E}$), in turn, sends the pair of entangled quantons (C$^\prime$ and B) to Charlie and Bob, respectively. Now, Charlie has two pairs of entangled quantons: one pair shared with Alice and the other with Bob. Subsequently, Charlie performs a Bell-basis measurement (BBM) on the two qubits in his possession. As a result, Alice and Bob---who initially shared no entanglement---end up with a maximally entangled pair of quantons. The result of the Charlie measurement can be shared by the classical communication channel to allow $\mathcal{A}$ and $\mathcal{B}$ to choose the specific entangled state. Although ESP is typically formulated in terms of qubits, recent advances in quantum information and computation involving qudits have motivated the extension of such a protocol to higher-dimensional quantum systems~\cite{Bouda2001, Bergou2021}.

In classical computation, bits are prevalent. However, studies have shown that the use of multivalued logic (with more than two levels) may offer advantages~\cite{Hurst1984, Bormashenko2019, Sandhie2021, Zhu2025}.  In quantum computation, qubits are widely used as the fundamental units of quantum information. Nonetheless, qudits have emerged as promising alternatives, offering several theoretical and practical benefits~\cite{Balantekin2024, Farias2025}. In particular, qudits enable the more efficient synthesis of arbitrary unitary operations, often requiring fewer entangling gates compared to qubit-based architectures~\cite{Di2015}. They also support the implementation of quantum algorithms with improved efficiency~\cite{Nikolaeva2024, Wang2020}, and can significantly reduce circuit complexity in various computational tasks~\cite{Lanyon2009}. Furthermore, qudits have been shown to enhance the performance of quantum error correction codes and fault-tolerant schemes~\cite{Anwar2012, Campbell2012, Campbell2014, Duclos-Cianci2013, Anwar2014, Andrist2015, Watson2015, Chizzini2022}, as well as to facilitate the design of optimal quantum measurement strategies~\cite{Stricker2022, Fischer2022}. In addition, they allow for richer structures in the study of entanglement complexity, potentially expanding the computational power and resource efficiency of quantum systems~\cite{Weggemans2022}.

Beyond their theoretical advantages, qudits have also seen increasing adoption in experimental platforms. Several physical implementations now support qudits, including trapped ions~\cite{Ringbauer2022, Hrmo2023}, photonic systems~\cite{Lu2020, Chi2022}, and Rydberg atoms~\cite{Ahn2000, Cohen2021, Weggemans2022, Gonzalez-Cuadra2022}, which naturally provide access to high-dimensional Hilbert spaces. Superconducting circuits have likewise demonstrated the feasibility of qudit-based processors by exploiting higher energy levels of transmon devices~\cite{Morvan2021, Fischer2023}. Other atomic systems offer distinct mechanisms for implementing qudits, further contributing to the diversity of scalable quantum platforms~\cite{Smith2013, Anderson2015, Senko2015, Kasper2021, Deller2023}.

The versatility of qudits has led to numerous uses in various areas of quantum science and technology. In quantum chemistry, they improve the encoding of electronic structure problems and decrease resource demands in variational algorithms~\cite{Cao2019, McArdle2020, MacDonell2021, Maskara2025}. In condensed matter physics, qudit-based models have facilitated the simulation of spin chains and correlated systems with higher spin representations~\cite{Haldane1983, Wecker2015, Sawaya2020}. Moreover, qudits have been proposed as effective encodings to solve combinatorial and optimization problems~\cite{Deller2023}, and to implement lattice gauge theories where nonbinary degrees of freedom are naturally present~\cite{Rico2018, Meth2025}. These applications illustrate the broad potential of qudits to extend the reach of quantum computation beyond what is achievable with qubits alone.

To fully characterize a quanton, it is necessary to resort to complete complementarity relations (CCRs), also known as triality relations. Rather than considering only the wave-particle features---known as complementarity relations (CRs)---CCRs also take into account correlations with other quantum systems. Recently, CCRs were obtained and explored in Refs.~\cite{Basso2020a, Basso2021, Basso2022b}, extending the work of Jakob and Bergou in Ref.~\cite{Jakob2010}, without resorting to retro-inference methods and focusing on a state-dependent approach~(see the discussion about retro-inference in Ref.~\cite{Araujo2025}). This result builds upon previous work in Ref.~\cite{Basso2020b}, which showed that CRs arise directly from the axioms of quantum mechanics, leading to an updated version of the quantum complementarity principle~\cite{Starke2024}.

\begin{figure}
    \centering
    \includegraphics[width=1\linewidth]{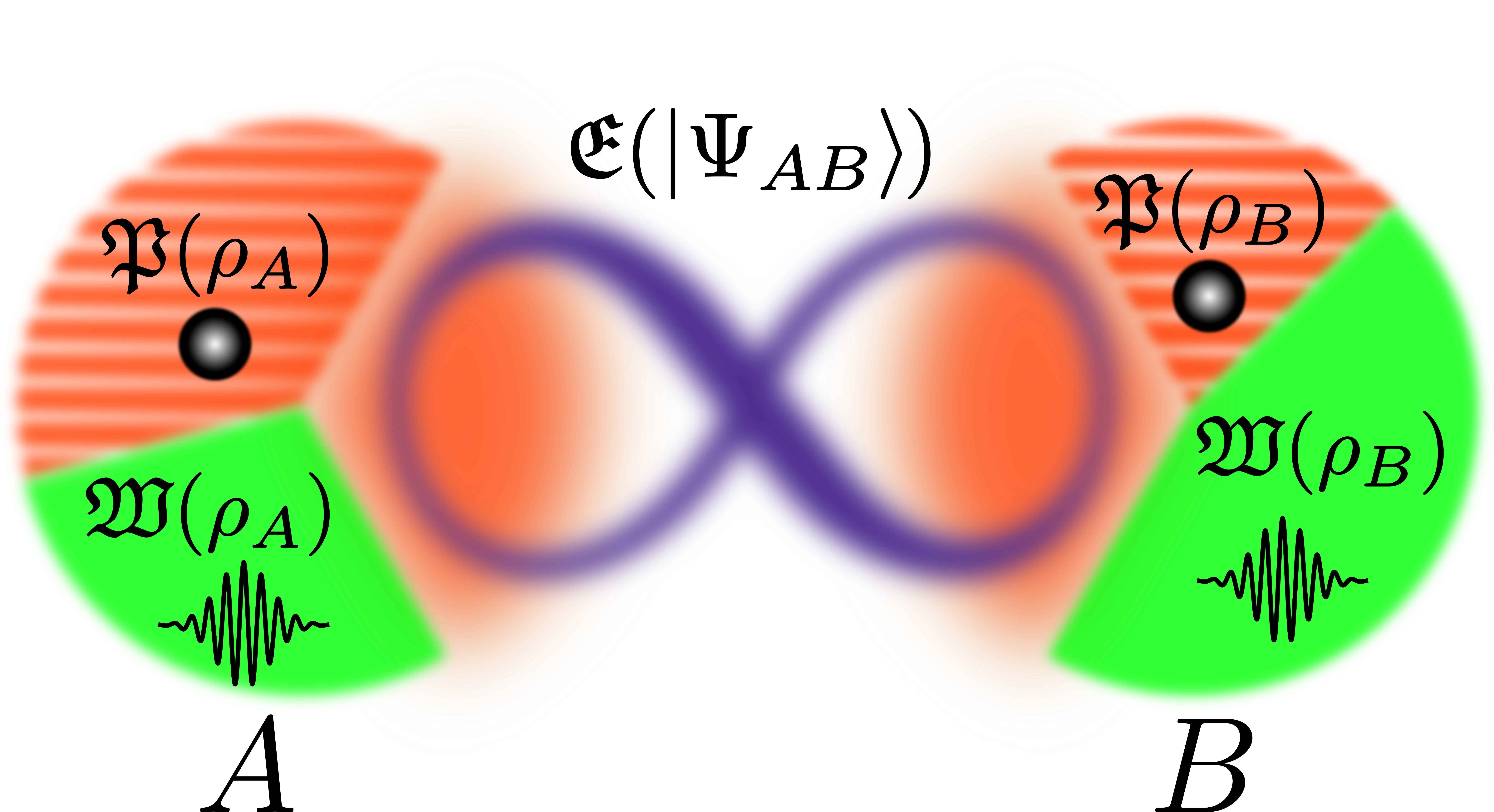}
    \caption{Diagram illustrating a broad scenario involving a pure entangled state shared between the quantons $A$ and $B$. The wave-particle features of a quanton are fully characterized by a complete complementarity relation, expressed in Eq.~\eqref{eq:CCR}, where $\mf{W}$ is the wave measure represented by quantum coherence, $\mf{P}$ is the predictability measure, and $\mf{E}$ is an entanglement monotone. The sum of these three quantities is equal to a constant $\alpha(d_A)$, which depends solely on the system's dimension. In general, the local quantities $\mf{P}$ and $\mf{W}$ may differ between the quantons, however the global quantity $\mf{E}(|\Psi_{AB}\rangle)$ is the same for both systems $A$ and $B$.
    }
    \label{fig:CCR}
\end{figure}

In general, CCRs can be written as
\begin{align}
\mf{W}(\rho_{A}) + \mf{P}(\rho_{A}) + \mf{E}(\ket{\Psi_{AB}}) = \alpha (d_{A}),
\label{eq:CCR}
\end{align}
where $\rho_{A} = \Tr_{B}{\ketbra{\Psi_{AB}}}$ is the reduced density matrix of the bipartite quantum state $\ket{\Psi_{AB}}$, $\mf{W}$ is the wave measure represented by quantum coherence, $\mf{P}$ is the predictability measure, $\mf{E}$ is an entanglement monotone, with the CCR being limited by a constant $\alpha (d_{A})$ that depends on the dimension of the quanton $A$. A similar expression holds for the quanton $B$.  Figure~\ref{fig:CCR} presents a diagram that illustrates a typical scenario to quantify $\mf{W}$ and $\mf{P}$ in relation to a certain $\mf{E}$ involving two entangled quantons, $A$ and $B$.
%
%
A CCR of this type can be exemplified by
\begin{equation}
C_{l_1}(\rho_{A}) + P_{l_1}(\rho_{A}) + E_{l_1}(\ket{\Psi_{AB}}) = d_{A} - 1,
\label{eq:ccr_l1}
\end{equation}
where $C_{l_1}(\rho_{A}) := \sum_{j\neq k} \left\vert{\rho_{jk}^A}\right\vert$ denotes the quantum coherence with the $l_1$-norm. The term $P_{l_1}(\rho_{A}) = d_A - 1 - \sum_{j\neq k} \sqrt{\rho_{jj}^A\rho_{kk}^A}$ quantifies the corresponding $l_1$-norm predictability. Finally, the $l_1$-norm of entanglement reads $E_{l_1}(\ket{\Psi_{AB}}) = \sum_{j\neq k} \left( \sqrt{\rho_{jj}^A\rho_{kk}^A} - \left\vert{\rho_{jk}^A}\right\vert \right)$.

Using the CCR formalism, Ref.~\cite{Basso2022a} shows that, starting from null local quantum coherence, the ESP with partially entangled pure qubit states allows Alice and Bob to obtain maximally entangled states with a small but non-zero probability---even when the initial entanglement is nearly absent. This result was extended in Ref.~\cite{Maziero2023}, where the authors considered a pair of qubits in an arbitrary pure quantum state and showed that, when the ESP yields two pairs of maximally entangled qubits, the pre-measurement local coherence and/or predictability is consumed and transformed into post-measurement entanglement. These works provide an important connection between the operationalization of CCRs and ESP, providing valuable insights for quantum communication protocols and quantum algorithms. The present work builds upon these advances in ESP by extending the results of Ref.~\cite{Basso2022a, Maziero2023} to the case of qudits.

%
%
The remainder of this article is organized as follows. In Sec.~\ref{sec:qudits}, we introduce the generalized Bell states and review the algebra of qudits. In Sec.~\ref{sec:nullcoherence}, we show that it is preferable to choose initial partially entangled states with vanishing local quantum coherence, based solely on the analysis of the CCR. We then extend the results of Refs.~\cite{Basso2022a, Maziero2023} to qudits in Sec.~\ref{sec:ESPPqudits}, and also discuss the qubit and qutrit cases. Finally, we present our conclusions in Sec.~\ref{sec:conclusion}.

%
%
\section{Generalized Bell states}
\label{sec:qudits}
In this section, we introduce the generalized Bell states, that is, the extension of the Bell states to qudits~\cite{bennett93, Karimipour2002, Sych2009}, along with the quantum gates required to generate them~\cite{Karimipour2002, Camps2021}.

The Fourier gate for qudits can be defined as
\begin{align}
F = \frac{1}{\sqrt{d}}\sum_{j,k=0}^{d-1}\omega^{jk}|j\rangle\langle k|,
\end{align}
which $F|k\rangle = \frac{1}{\sqrt{d}}\sum_{j=0}^{d-1}\omega^{jk}|j\rangle$ is the action on a computational basis, where
\begin{align}
\omega = e^{2\pi i/d}, \label{eq:omega}   
\end{align}
and $\bar{\omega}$ denotes its complex conjugate. The CNOT gate for qudits can be constructed as follows. Given the state shift operator
\begin{align}
X(j) = \sum_{k=0}^{d-1} \ketbra{j\oplus k}{k},
\end{align}
in the computational basis, the gate operates as $X(j)\ket{k} = \ket{j\oplus k}$, with $j\oplus k = (j+k) \bmod d$  being the addition modulo $d$. Then, the CNOT gate can be defined as
\begin{align}
C_{X(j)}^{A\rightarrow B} = \sum_{j=0}^{d-1}|j\rangle_A\langle j|\otimes X(j)_B,
\end{align}
which, when operating on a computational basis, perform as $C_{X(j)}^{A\rightarrow B}|jk\rangle_{AB} = |j\rangle_A \otimes|j\oplus k\rangle_B$, where in the subscript $A\rightarrow B$ the control qudit is denoted as $A$, while $B$ represents the target qudit.

From these quantum gates, we can define the generalized Bell states as
\begin{align}
\begin{aligned}
|\Phi_{pq}^{AB}\rangle & =  C_{X(j)}^{B\rightarrow A}(\mathbb{I}_A \otimes F_B)|pq\rangle_{AB}   \\
&  = \frac{1}{\sqrt{d}}\sum_{j=0}^{d-1}\omega^{jq}|p\oplus j\rangle_A\otimes|j\rangle_B. \label{eq:gbellstates}
\end{aligned}
\end{align}

Thus, to prepare such states, we consider two qudits initialized in the state $\ket{0}$. We then apply the Fourier gate---a generalization of the Hadamard gate for qudits---to one of the qudits, followed by a CNOT gate acting on both. Moreover, we refer to the set of generalized Bell states given by Eq.~\eqref{eq:gbellstates} as the Bell basis, which is orthonormal and complete.

It is instructive to recover the standard Bell states for qubits by setting $d = 2$ in the qudit algebra.
From Eqs. \eqref{eq:omega} and~\eqref{eq:gbellstates}, the standard Bell states are obtained as follows
\begin{align}
\begin{aligned}
|\Phi_{pq}^{AB}\rangle 
& = \frac{1}{\sqrt{2}}\sum_{j=0}^1 e^{2\pi ijq/2}|(p+j)\bmod 2\rangle_A\otimes|j\rangle_B \\
& = \frac{1}{\sqrt{2}}\big(|p\rangle_A\otimes|0\rangle_B + e^{\pi iq}|p\oplus 1\rangle_A\otimes|1\rangle_B\big).
\end{aligned}
\end{align}
Hence, we recover the well-known Bell basis
\begin{align}
|\Phi_{00}\rangle & = \big(|0\rangle\otimes|0\rangle + e^{\pi i0}|1\rangle\otimes|1\rangle\big) /\sqrt{2} \equiv |\Phi_+\rangle, \\
|\Phi_{01}\rangle & = \big(|0\rangle\otimes|0\rangle + e^{\pi i}|1\rangle\otimes|1\rangle\big) /\sqrt{2} \equiv |\Phi_-\rangle, \\
|\Phi_{10}\rangle & = \big(|1\rangle\otimes|0\rangle + e^{\pi i0}|0\rangle\otimes|1\rangle\big) /\sqrt{2} \equiv |\Psi_+\rangle, \\
|\Phi_{11}\rangle & = \big(|1\rangle\otimes|0\rangle + e^{\pi i}|0\rangle\otimes|1\rangle\big) /\sqrt{2} \equiv -|\Psi_-\rangle.
\end{align}

%
%
\section{Zero local coherence states are enough}
\label{sec:nullcoherence}
In Ref.~\cite{Maziero2023}, the authors examined arbitrary pure two-qubit states and demonstrated that, in scenarios where ESP produces two maximally entangled qubit pairs, the initial local quantum resources---specifically coherence and/or predictability---are consumed and converted into entanglement after the measurement. In this section, we argue that, since local unitary transformations do not affect entanglement, one can employ initial partially entangled states with vanishing local coherence.

Let us assume that the initial states that Charlie shared with Alice and Bob are general pure states of the type
\begin{align}
& |\xi\rangle_{AC} = \sum_{j,k=0}^{d-1}c_{jk}|jk\rangle_{AC}, \\ 
& |\eta\rangle_{C'B} = \sum_{j,k=0}^{d-1}d_{jk}|jk\rangle_{C'B}.
\end{align}
In what follows, our analysis focuses on the quantum system shared between Alice and Charlie. Nevertheless, a similar line of reasoning can be applied to the quantum system shared between Bob and Charlie. From Eq.~\eqref{eq:CCR}, the purity of the first global quantum state allows us to express the CCR as
\begin{align}
\begin{aligned}
C_{l_1}(\rho_{\xi_A}) + P_{l_1}(\rho_{\xi_A}) + E_{l_1}(\ket{\xi}_{AC}) = d - 1,\\
C_{l_1}(\rho_{\xi_C}) + P_{l_1}(\rho_{\xi_C}) + E_{l_1}(\ket{\xi}_{AC}) = d - 1,
\end{aligned}
\end{align}
where $\rho_{\xi_A} = \Tr_{C}(\ketbra{\xi_{AC}})$ and $\rho_{\xi_C} = \Tr_{A}(\ketbra{\xi_{AC}})$, with the assumption that all qudits have the same dimension.

As local unitary transformations do not change the entanglement shared between the subsystems $A$ and $C$, the states $|\xi'\rangle_{AC}$ and $|\xi\rangle_{AC}$ connected by the local unitary $U_l$, i.e., $|\xi'\rangle_{AC} = U_l|\xi\rangle_{AC}$, are such that:
\begin{align}
    &C_{l_1}(\rho_{\xi'_A}) + P_{l_1}(\rho_{\xi'_A})  = C_{l_1}(\rho_{\xi_A}) + P_{l_1}(\rho_{\xi_A}), \\ 
    & C_{l_1}(\rho_{\xi'_C}) + P_{l_1}(\rho_{\xi'_C}) = C_{l_1}(\rho_{\xi_C}) + P_{l_1}(\rho_{\xi_C}).
\end{align}
This, in turn, implies that we can choose a basis in which the reduced states $\rho_{\xi_A}$ and $\rho_{\xi_C}$ are diagonal. Consequently, the local quantum coherence of each subsystem vanishes, i.e., $C_{l_1}(\rho_{\xi_A}) = C_{l_1}(\rho_{\xi_C}) =0$. In this case, the CCR simplifies to
\begin{align}
 E_{l_1}(|\xi\rangle_{AC}) = d-1 - P_{l_1}(\rho_{\xi_A}) = d-1 - P_{l_1}(\rho_{\xi_C}).
\end{align}

Thus, without loss of generality, we may utilize bipartite quantum states with null local quantum coherence as in Ref.~\cite{Basso2022a}, which are given by
\begin{align}
& |\xi\rangle_{AC} = \sum_{j=0}^{d-1}c_{j}|jj\rangle_{AC},\\ & |\eta\rangle_{C'B} = \sum_{k=0}^{d-1}d_{k}|kk\rangle_{C'B},
\end{align}
with $\sum_{j= 0}^{d-1} |c_j|^2 = 1$ and $\sum_{j= 0}^{d-1} |d_j|^2 = 1$.

%
%
\section{Entanglement swapping protocol for partially entangled qudits}
\label{sec:ESPPqudits}

In this section, we present the ESP for partially entangled qudits, extending the results reported in Refs. in~\cite{Basso2022a, Maziero2023}, and we derive our main results.

We begin by considering the laboratories operated by Alice, Bob, Charlie, Darwin, and Erin, as depicted in Fig.~\ref{fig:ESwap}. Darwin prepares a partially entangled pair of qudits (A and C) and sends them to Alice and Charlie, respectively. Erin also prepares a pair of partially entangled qudits (C$^\prime$ and B) and sends them to Charlie and Bob, respectively. The initial composed state of the four qudits can be written as
\begin{align}
\begin{aligned}
|\Psi\rangle & =  \sum_{j=0}^{d-1}c_{j}|jj\rangle_{AC}\otimes \sum_{k=0}^{d-1}d_{k}|kk\rangle_{C'D} \\ & = \sum_{j,k=0}^{d-1}c_{j}d_{k}|jk\rangle_{AB}\otimes|jk\rangle_{CC'},
\label{eq:initialpsi}
\end{aligned}
\end{align}
with reduced states being given by
\begin{align}
\rho_{\xi_A} &=\rho_{\xi_C}=\sum_{j=0}^{d-1}|c_j|^2|j\rangle\langle j|, \\
\rho_{\eta_{C'}} &=\rho_{\eta_B}=\sum_{k=0}^{d-1}|d_k|^2|k\rangle\langle k|.
\end{align}
The local quantum coherence vanishes for all reduced states, i.e., $C_{l_1}(\rho_s)=0$ for $s=A,B,C,C'$, and the initial entanglement and predictability of the reduced states are given by
\begin{align}
& E_{l_1}(|\xi\rangle_{AC}) = \sum_{j\ne k}|c_j c_k|, \\
& E_{l_1}(|\eta\rangle_{C'B}) = \sum_{j\ne k}|d_j d_k|, \\
& P_{l_1}(\rho_{\xi_A}) = P_{l_1}(\rho_{\xi_C}) = d-1-\sum_{j\ne k}|c_j c_k|, \\ 
& P_{l_1}(\rho_{\eta_{C'}}) = P_{l_1}(\rho_{\eta_B}) = d-1-\sum_{j\ne k}|d_j d_k|.
\end{align}

After Charlie makes a BBM, the resulting non-normalized state is
\begin{align}
\begin{aligned}
\ket{\phi_{pq}} & = \big(\mathbb{I}_{AB}\otimes \big| \Phi_{pq}^{CC'} \big\rangle \big \langle \Phi_{pq}^{CC'} \big|\big)\ket{\Psi} \\
& = \sum_{j,k=0}^{d-1}c_j d_k \frac{1}{\sqrt{d}}\bar{\omega}^{qk}\delta_{j,p\oplus k} |jk\rangle_{AB} \big|\Phi_{pq}^{CC'}\big\rangle \\
& = \Big(\frac{1}{\sqrt{d}}\sum_{k=0}^{d-1}c_{p\oplus k} d_k \bar{\omega}^{qk}|p\oplus k,k\rangle_{AB}\Big) \big|\Phi_{pq}^{CC'}\big\rangle,
\end{aligned}
\end{align}
where we used
\begin{align}
\begin{aligned}
\langle\Phi_{pq}|jk\rangle & = \left(\frac{1}{\sqrt{d}}\sum_{l=0}^{d-1}\bar{\omega}^{ql}\langle p\oplus l|\otimes\langle l|\right)|jk\rangle \\
& = \frac{1}{\sqrt{d}}\sum_{l=0}^{d-1}\bar{\omega}^{ql}\langle p\oplus l|j\rangle\otimes\langle l|k\rangle \\
& = \frac{1}{\sqrt{d}}\bar{\omega}^{qk}\delta_{j,p\oplus k}.
\end{aligned}
\end{align}
For Alice and Bob, the post-BBM non-normalized state is
\begin{align}
    \ket{\phi_{pq}^{AB}} = \frac{1}{\sqrt{d}}\sum_{k=0}^{d-1}c_{p\oplus k} d_k \bar{\omega}^{qk}|p\oplus k,k\rangle.
\end{align}
To normalize the post-BBM state, let us notice that
\begin{align}
& \left\Vert\ket{\phi_{pq}^{AB}}\right\Vert^2 = \braket{\phi_{pq}^{AB}} \nonumber \\
& = \frac{1}{d}\sum_{k,l=0}^{d-1}c_{p\oplus k} d_k \bar{c}_{p\oplus l} \bar{d}_l \omega^{k-l}\langle p\oplus k,k|p\oplus l,l\rangle \\
& = \frac{1}{d}\sum_{k=0}^{d-1}|c_{p\oplus k}|^2 |d_k|^2 = \Pr\big(\Phi_{pq}^{CC'}\big), \nonumber 
\end{align}
where $\Pr\big(\Phi_{pq}^{CC'}\big)$ is the probability that Charlie will obtain the Bell state $\big|\Phi_{pq}^{CC'}\big\rangle$ in the BBM, with
\begin{align}
&\sum_{p = 0}^{d-1} \Pr\big(\Phi_{pq}^{CC'}\big) = \frac{1}{d}\sum_{j=0}^{d-1}|d_j|^2 \left(\sum_{p = 0}^{d-1} |c_{p\oplus j}|^2 \right) \\
&= \frac{1}{d}\sum_{j=0}^{d-1}|d_j|^2 = \frac{1}{d},
\end{align} 
and therefore
$ \sum_{p,q = 0}^{d-1} \Pr\big(\Phi_{pq}^{CC'}\big) = 1.$

Hence, we can write the initial state given by Eq.~\eqref{eq:initialpsi} in terms of the post-BBM state, i.e.,
\begin{align}
\begin{aligned}
|\Psi\rangle & = \sum_{p,q = 0}^{d-1} \left\Vert\ket{\phi_{pq}^{AB}}\right\Vert\frac{\ket{\phi_{pq}^{AB}}}{\left\Vert\ket{\phi_{pq}^{AB}}\right\Vert} \otimes \big|\Phi_{pq}^{CC'}\big\rangle \\
& = \sum_{p,q = 0}^{d-1} \sqrt{\Pr\left(\Phi_{pq}^{CC'}\right)} \big|\hat{\phi}_{pq}^{AB}\big\rangle \otimes \big|\Phi_{pq}^{CC'}\big\rangle,
\end{aligned}
\end{align}
where
\begin{align}
\begin{aligned}
\ket{\hat{\phi}_{pq}^{AB}} & = \frac{\ket{\phi_{pq}^{AB}}}{\left\Vert\ket{\phi_{pq}^{AB}}\right\Vert} = \sum_{k=0}^{d-1}\frac{c_{p\oplus k} d_k \omega^{qk}}{\sqrt{ \Pr\left(\Phi_{pq}^{CC'}\right) d}}|p\oplus k\rangle_A\otimes|k\rangle_B,
\end{aligned}
\end{align}
is the normalized quantum state of Alice and Bob.

From the reduced state of Alice (or Bob), which is given by
\begin{align}
\begin{aligned}
\rho_{\hat{\phi}_{pq}^{A}} & = \Tr_B \big(\big|\hat{\phi}_{pq}^{AB}\big\rangle \big\langle\hat{\phi}_{pq}^{AB}\big| \big) \\
& = \frac{1}{\Pr\left(\Phi_{pq}^{CC'}\right) d}\sum_{k=0}^{d-1}c_{p\oplus k}^2 d_k^2|p\oplus k\rangle\langle p\oplus k|,
\end{aligned}
\end{align}
the amount of entanglement contained in the post-BBM state of Alice and Bob is given by
\begin{align}
\begin{aligned}
E_{l_1} \big( |\hat{\phi}_{pq}^{AB} \rangle \big) & = \sum_{j\ne k}\sqrt{\left(\rho_{\hat{\phi}_{pq}^{A}}\right)_{jj} \left(\rho_{\hat{\phi}_{pq}^{A}}\right)_{kk}} \\
& = \frac{1}{\Pr\left(\Phi_{pq}^{CC'}\right) d}\sum_{j\ne k} \left\vert c_{p\oplus j} c_{p\oplus k} d_j d_k\right\vert. \label{eq:EphiAB}
\end{aligned}
\end{align}
From Eq.~\eqref{eq:EphiAB}, we can see that if the initial local predictabilities are maximal (zero initial entanglement), i.e., $c_j=1,\ d_j=1$ for some $j,$ then all other coefficients vanish, and, consequently, the final entanglement is zero. If the initial predictabilities are zero (maximum initial entanglement), i.e., $c_j = d_j=\frac{1}{\sqrt{d}}$ for all $j$, then the final entanglement shared between Alice and Bob reaches its maximum:
\begin{align}
E_{l_1} \big(|\hat{\phi}_{pq}^{AB}\rangle \big) = d-1,
\end{align}
with Charlie's corresponding probability given by
\begin{align}
\Pr\big(\Phi_{pq}^{CC'}\big) =  \frac{1}{d^2}.
\end{align}
These results are consistent with what one expects for these particular cases.

Now, we observe that the following inequality holds
\begin{align}
& \sum_{p = 0}^{d-1} \sum_{j\ne k}|c_{p\oplus j} c_{p\oplus k} d_j d_k| = \sum_{j\ne k}\!|d_j d_k| \! \left(\sum_{p = 0}^{d-1} |c_{p\oplus j} c_{p\oplus k}|\right) \nonumber \\
& \le  \sum_{j\ne k}|d_j d_k| \frac{1}{2} \sum_{p = 0}^{d-1} (|c_{p\oplus j}|^2 + |c_{p\oplus k}|^2) =  \sum_{j\ne k}|d_j d_k| \nonumber \\ 
& = E_{l_1}(|\eta\rangle_{C'B}), \label{eq:inequality} 
\end{align}
where the inequality follows from the elementary bound $(|a| - |b|)^2 \ge 0$ and we use the normalization condition $\sum_{p = 0}^{d-1} |c_{p\oplus j}|^2 = 1$. 

Combining Eq.~\eqref{eq:EphiAB} with the inequality~\eqref{eq:inequality}, we obtain
\begin{align}
\begin{aligned}
\big\langle E_{l_1}\big(|\hat{\phi}_{pq}^{AB} \rangle\big) \big\rangle  & = \sum_{p,q = 0}^{d-1} \Pr\big(\Phi_{pq}^{CC'}\big) E_{l_1}\big( |\hat{\phi}_{pq}^{AB} \rangle \big) \\ & = \sum_{q = 0}^{d-1} \frac{1}{d} \sum_{p = 0}^{d-1} \sum_{j\ne k}|c_{p\oplus j} c_{p\oplus k} d_j d_k| \\ &  \le E_{l_1}(|\eta\rangle_{C'B})
\\ & = d-1 - P_{l_1}(\rho_{\eta_B}).
\label{eq:average}
\end{aligned}
\end{align}
This is an interesting result that complements the findings of Refs.~\cite{Basso2022a, Maziero2023}, where it was shown that Alice and Bob can end up with maximally entangled states with a small but non-zero probability, even when starting from partially entangled qubits. On the other hand, Eq.~\eqref{eq:inequality} shows that, on average, the entanglement shared between Alice and Bob cannot exceed the initial entanglement of one of the pairs. Owing to the constraints imposed by the CCRs, the average entanglement is fully determined by the initial local predictability of subsystem $B$.

It is worth mentioning that this inequality holds for qudits in general and the upper bound is saturated when the initial qudits are maximally entangled. Moreover, a similar upper bound exists in terms of $E_{l_1}(|\xi\rangle_{AC})$.

In addition, the following inequality also holds
\begin{align}
\begin{aligned}
\hspace{-0.2cm}\sum_{p = 0}^{d-1} \sum_{j\ne k}|c_{p\oplus j} c_{p\oplus k} d_j d_k| & \le  \sum_{j\ne k}|d_j d_k| \sum_{l\neq m} |c_l c_m|,  \\
& = E_{l_1}(|\xi\rangle_{AC}) E_{l_1}(|\eta\rangle_{C'B}). \label{eq:inequality1} 
\end{aligned}
\end{align}
To understand the reasoning behind this inequality, let us note that, by fixing $j$ and $k$ such that $j \neq k$, it implies that $p\oplus j \neq p \oplus k$. Defining $l = p\oplus j$ and $m = p \oplus k $, then $l \neq m$. However, the sum over $p$ contains $d$ positive terms, while the sum over $l \neq m$ contains $d(d-1)$ positive terms.

Combining Eq.~\eqref{eq:EphiAB} with the inequality~\eqref{eq:inequality1}, we obtain
\begin{align}
\begin{aligned}
\big\langle E_{l_1}\big(|\hat{\phi}_{pq}^{AB} \rangle\big) \big\rangle  & = \sum_{p,q = 0}^{d-1} \Pr\big(\Phi_{pq}^{CC'}\big) E_{l_1}\big( |\hat{\phi}_{pq}^{AB} \rangle \big) \\ &  \le E_{l_1}(|\xi\rangle_{AC}) E_{l_1}(|\eta\rangle_{C'B}) \\
& \hspace{-1.2cm} = \left(d - 1 - P_{l_1}(\rho_{\xi_A}) \right) \left(d - 1 - P_{l_1}(\rho_{\eta_B}) \right),
\label{eq:average1}
\end{aligned}
\end{align}
Thus, the average distributed entanglement is upper bounded by the product of the initial entanglements, which, due to the constraints imposed by the CCRs, are entirely determined by the initial local predictabilities of subsystems $A$ and $B$. In particular, the greater the local predictability, the lower the average distributed entanglement.

Next, we consider the effect of post-selection in the BBM and obtain an upper bound for the entanglement of each resulting post-measurement state of qudits A and B in terms of their initial entanglements with $C$ and $C'$, respectively, and the probability that Charlie will obtain the Bell state $\ket{\Phi_{pq}^{CC'}}$ after the BBM. From Eqs.~\eqref{eq:EphiAB} and~\eqref{eq:inequality1}, it follows that
\begin{align}
E_{l_1}\big( |\hat{\phi}_{pq}^{AB}\rangle \big) \le \frac{E_{l_1}(|\eta\rangle_{C'B}) E_{l_1}(|\eta\rangle_{C'B})}{\Pr(|\Phi_{pq}\rangle_{CC'}) d}. \label{eq:upperbound}
\end{align}
By using the CCRs, it is easy to see that we can express the upper bound given by Eq.~\eqref{eq:upperbound} in terms of the initial local predictability of $A$ and $B$. 

\subsection{The qubit case}

\begin{figure*}
    \centering
    \includegraphics[width=1\linewidth]{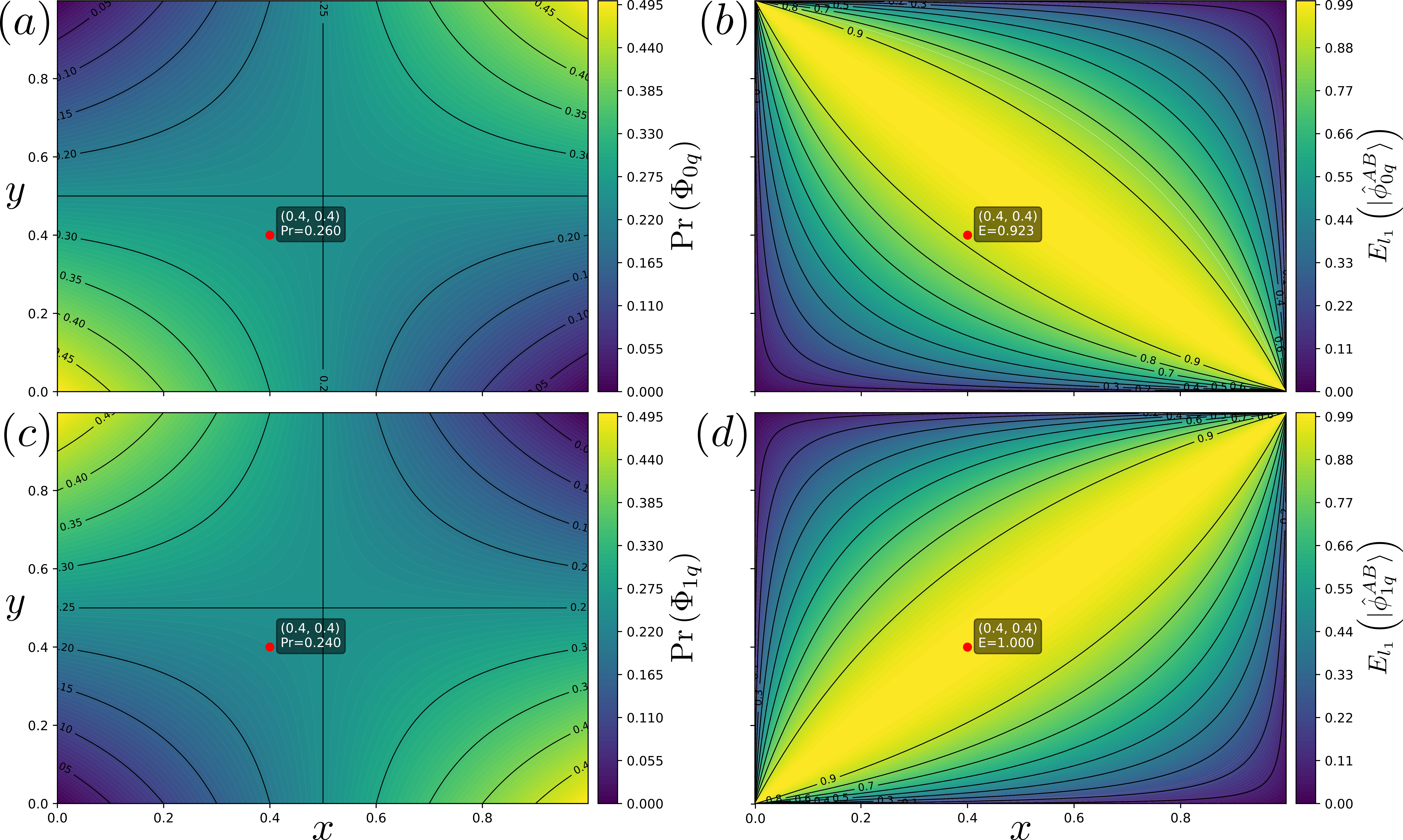}
    \caption{The probabilities $\Pr\big(\Phi_{0q}^{CC'}\big)$ and $\Pr\big(\Phi_{1q}^{CC'}\big)$, together with the post-BBM entanglements $E_{l_1} \big( |\hat{\phi}_{0q}^{AB}\rangle \big)$ and $E_{l_1} \big( |\hat{\phi}_{1q}^{AB}\rangle \big)$, as functions of the parameters $x$ and $y$ for the initial states given by Eq.~\eqref{eq:initqubits}.}
    \label{fig:qubitcase}
\end{figure*}

Considering the particular case of qubits, i.e., $d=2$, we recover the results obtained in Ref.~\cite{Maziero2023}. Interestingly, we will see that, in this case, the upper bounds given by Eqs.~\eqref{eq:average1} and~\eqref{eq:upperbound} are saturated.

Let us start by noticing that for $p = 0$, we shall have
\begin{align}
\begin{aligned}
E_{l_1} \big(|\hat{\phi}_{0q}^{AB}\rangle \big) = \frac{|c_0 c_1 d_0 d_1|}{\Pr(\Phi_{0q}^{CC'})},
\end{aligned}
\end{align}
and
\begin{align}
\begin{aligned}
E_{l_1}(|\xi\rangle_{AC})E_{l_1}(|\eta\rangle_{C'B}) = 4|c_0 c_1 d_0 d_1|,
\end{aligned}
\end{align}
which implies that
\begin{equation}
E_{l_1} \big(|\hat{\phi}_{0q}^{AB}\rangle \big) = \frac{E_{l_1}(|\xi\rangle_{AC})E_{l_1}(|\eta\rangle_{C'B})}{4\Pr(\Phi_{0q}^{CC'})},\label{eq:qubit}
\end{equation}
which corresponds to the saturated upper bound given by Eq.~\eqref{eq:upperbound} for $d = 2$ and Bell states characterized by $p = 0$. The same result also holds for $p = 1$.

Finally, Eq.~\eqref{eq:average1} for the qubit case also follows from Eq.~\eqref{eq:qubit}. Noting that
\begin{align}
\sum_{q = 0}^{1}\Pr\big(\Phi_{0q}^{CC'}\big)E_{l_1} \big(|\hat{\phi}_{0q}^{AB}\rangle \big) & = \frac{1}{4}\sum_{q = 0}^{1} E_{l_1}(|\xi\rangle_{AC})E_{l_1}(|\eta\rangle_{C'B}) \nonumber \\ & = \frac{1}{2} E_{l_1}(|\xi\rangle_{AC})E_{l_1}(|\eta\rangle_{C'B}), \nonumber
\end{align}
and that this relation also holds for $p = 1$, it follows that
\begin{align}
\begin{aligned}
\big\langle E_{l_1}\big(|\hat{\phi}_{pq}^{AB}\rangle\big) \big\rangle  & = \sum_{p,q = 0}^{1} \Pr\big(\Phi_{pq}^{CC'}\big) E_{l_1}\big( |\hat{\phi}_{pq}^{AB} \rangle \big) \\ &  = E_{l_1}(|\xi\rangle_{AC}) E_{l_1}(|\eta\rangle_{C'B}). \label{eq:qubitE}
\end{aligned}
\end{align}
Hence, for qubits, the average distributed entanglement is exactly equal to the product of the initial entanglements. This result was obtained in Refs.~\cite{Bergou2021, Maziero2023} using the concurrence as the entanglement measure.

Now, let us consider, as an explicit example, the initial states analyzed in Ref.~\cite{Basso2022a}, namely
\begin{align}
\begin{aligned}
& |\xi\rangle_{AC} = \sqrt{x}|00\rangle_{AC} +  \sqrt{1-x}|11\rangle_{AC},\\ & |\eta\rangle_{C'B} = \sqrt{y}|00\rangle_{C'B} +  \sqrt{1-y}|11\rangle_{C'B}, \label{eq:initqubits}
\end{aligned}
\end{align}
with $x,y \in [0,1]$. In this case, the initial entanglement between each pair is given by
\begin{align}
& E_{l_1}(|\xi\rangle_{AC}) = 2\sqrt{x(1-x)}, \\
& E_{l_1}(|\eta\rangle_{C'B}) = 2\sqrt{y(1-y)}, 
\end{align}
while Charlie's probabilities of obtaining a Bell state after the BBM are
\begin{align}
&\Pr\big(\Phi_{0q}^{CC'}\big) = \frac{1}{2}\left(xy + (1-x)(1-y)\right), \\
&\Pr\big(\Phi_{1q}^{CC'}\big) = \frac{1}{2}\left(x(1-y) + (1-x)y\right),
\end{align}
with the post-BBM entanglement between Alice and Bob being given by
\begin{align}
& E_{l_1} \big( |\hat{\phi}_{0q}^{AB}\rangle\big) =\frac{xy(1-x)(1-y)}{xy + (1-x)(1-y)}, \\
& E_{l_1} \big( |\hat{\phi}_{1q}^{AB} \rangle\big) =\frac{xy(1-x)(1-y)}{x(1-y) + (1-x)y}.
\end{align}

The probabilities $\Pr \big( \Phi_{0q}^{CC'} \big)$ and $\Pr \big(\Phi_{1q}^{CC'}\big)$, together with the post-BBM entanglements $E_{l_1} (|\hat{\phi}_{0q}^{AB}\rangle )$ and $E{l_1} \big( |\hat{\phi}_{1q}^{AB}\rangle \big)$, are shown in Fig.~\ref{fig:qubitcase} as functions of the parameters $x$ and $y$. For $x = y = 1/2$, the probabilities $\Pr\big(\Phi_{0q}^{CC'}\big)$ and $\Pr\big(\Phi_{1q}^{CC'}\big)$ are equal to $1/2$, and the post-BBM entanglements $E_{l_1} \big( |\hat{\phi}_{0q}^{AB}\rangle \big)$ and $E_{l_1} \big( |\hat{\phi}_{1q}^{AB}\rangle \big)$ are equal to unity, i.e., they are maximal, as expected. On the other hand, as one moves along the line $x = y$ towards $x \to 0$ or $x \to 1$, the probability $\Pr\big(\Phi_{0q}^{CC'}\big)$ tends to $1/2$, while the post-BBM entanglement $E_{l_1} \big( |\hat{\phi}_{0q}^{AB}\rangle \big)$ tends to zero. Conversely, the probability $\Pr\big(\Phi_{1q}^{CC'}\big)$ tends to zero, and the entanglement $E_{l_1} \big( |\hat{\phi}_{1q}^{AB}\rangle \big)$ remains equal to $1$ along the entire line $x = y$.

Moreover, for the point $x = y = 0.4$, which is depicted as a red dot in each panel of Fig.~\ref{fig:qubitcase}, we have that the probabilities are $\Pr\big(\Phi_{0q}^{CC'}\big) = 0.26$ and $\Pr\big(\Phi_{1q}^{CC'}\big) = 0.24$, with the corresponding post-BBM entanglements $E_{l_1} \big( |\hat{\phi}_{0q}^{AB}\rangle \big) = 0.923$ and $E_{l_1} \big( |\hat{\phi}_{1q}^{AB}\rangle \big) = 1$. The initial entanglements are $E_{l_1}(|\xi\rangle_{AC}) = E_{l_1}(|\eta\rangle_{C'B}) = 0.9797$. Therefore, depending on the outcome of Charlie's BBM, the entanglement between Alice and Bob's pair of qubits can increase to unity. From the CCR perspective, this implies that the predictability of the initial states was consumed in order to increase the entanglement. Finally, it is interesting to notice that the average post-BBM entanglement is $\big\langle E_{l_1}\big(|\hat{\phi}_{pq}^{AB} \rangle \big) \big\rangle = 0.95996$, which is less than the initial entanglement, as predicted by Eq.~\eqref{eq:average}.

\subsection{The qutrit case and a conjecture}

Here we explore the qutrit case, i.e., $d = 3$, and show that the average distributed entanglement turns out to be half the product of the initial entanglements.  This result suggests that the upper bound given by Eq.~\eqref{eq:average1} can be improved. Although we are not able to provide a proof, we conjecture what this improved upper bound might be.

Let us begin by noticing that
\begin{align}
\begin{aligned}
\sum_{p = 0}^{2} \sum_{j\ne k}|c_{p\oplus j} c_{p\oplus k} d_j d_k| &  = 2\Big(|d_0 d_1|  \sum_{p = 0}^{2} |c_{p\oplus 0} c_{p\oplus 1}| \\ & + |d_0 d_2| \sum_{p = 0}^{2} |c_{p\oplus 0} c_{p\oplus 2}| \\& + |d_1 d_2| \sum_{p = 0}^{2} |c_{p\oplus 1} c_{p\oplus 2}|\Big) \\ 
& = \frac{1}{2}E_{l_1}(|\xi\rangle_{AC}) E_{l_1}(|\eta\rangle_{C'B}),
\end{aligned}
\end{align}
which gives us the following average over all post-measurement entangled states
\begin{align}
\begin{aligned}
\big\langle E_{l_1} \big(|\hat{\phi}_{pq}^{AB}\rangle \big) \big\rangle &= \sum_{p,q = 0}^{2} \Pr\big(\Phi_{pq}^{CC'}\big) E_{l_1} \big(|\hat{\phi}_{pq}^{AB}\rangle \big) \\& = \sum_{q = 0}^{2} \frac{1}{3} \sum_{p = 0}^{2} \sum_{j\ne k}|c_{p\oplus j} c_{p\oplus k} d_j d_k| \\& = \frac{1}{2} E_{l_1}(|\xi\rangle_{AC}) E_{l_1}(|\eta\rangle_{C'B}), \label{eq:qutrit}
\end{aligned}
\end{align}
which satisfies the upper bound given by Eq.~\eqref{eq:average}, but does not saturate it.

Another point to note is that we choose not to normalize the entanglement measure $E_{l_1}$. If we do normalize, i.e., $E_{l_1} \to \mathcal{E}_{l_1} = E_{l_1}/(d-1)$, then Eq.~\eqref{eq:qutrit} can be recast as
\begin{align}
    \big\langle \mathcal{E}_{l_1} \big(|\hat{\phi}_{pq}^{AB}\rangle \big) \big\rangle = \mathcal{E}_{l_1}(|\xi\rangle_{AC}) \mathcal{E}_{l_1}(|\eta\rangle_{C'B}), \label{eq:normE}
\end{align}
exactly as in the qubit case.

The exact results given by Eqs.~\eqref{eq:qubitE} and~\eqref{eq:qutrit} suggest that the average entanglement is, at the very least, upper bounded by a factor that depends on the system's dimension times the product of the initial entanglements. Therefore, we conjecture that the improved upper bound has the form
\begin{align}
\big\langle E_{l_1}\big(|\hat{\phi}_{pq}^{AB} \rangle\big) \big\rangle    \le \frac{E_{l_1}(|\xi\rangle_{AC}) E_{l_1}(|\eta\rangle_{C'B})}{d-1}. 
\label{eq:conjecture}
\end{align}
This conjecture is motivated by the following facts. First, it is saturated for $d= 2$ and $d=3$, reproducing the exact results given by Eqs.~\eqref{eq:qubitE} and~\eqref{eq:qutrit}. Second, it is also saturated for maximally entangled qudits. Moreover, if we normalize the entanglement measure as $E_{l_1} \to \mathcal{E}_{l_1} = E_{l_1}/(d-1)$, the conjecture reduces to $ \big\langle \mathcal{E}_{l_1} \big(|\hat{\phi}_{pq}^{AB}\rangle \big)\big\rangle \le \mathcal{E}_{l_1}(|\xi\rangle_{AC}) \mathcal{E}_{l_1}(|\eta\rangle_{C'B})$, which is independent of the dimension. Finally, let us notice that the conjecture given by Eq.~\eqref{eq:conjecture} is equivalent to showing that, for fixed indices $j$ and $k$, $\sum_{p} |c_{p\oplus j} c_{p\oplus k}|  \le \sum_{l \neq m} | c_l c_m |/(d-1)$. Therefore, in Eq.~\eqref{eq:inequality1}, this corresponds to replacing the sum over $p$, which contains $d$ positive terms, with the sum over $l \neq m$, which contains $d(d-1)$ positive terms, weighted by a factor of $1/(d-1)$.

%
%
\section{Final Remarks}
\label{sec:conclusion}
In this work, we extended the entanglement swapping protocol (ESP) to partially entangled qudit states and investigated its behavior within the framework of complete complementarity relations. We demonstrated that initial states with vanishing local quantum coherence are sufficient to study the protocol, as any pure bipartite state can be locally transformed into a form with diagonal marginals without altering the entanglement content. This simplification allowed us to obtain analytical expressions for the distributed entanglement after a Bell-basis measurement and to identify relevant bounds.

We established two upper bounds for the average distributed entanglement between the remote parties: one based on the initial entanglement of a single pair, and another based on the product of the initial entanglements. These results generalize and complement previous findings for qubits~\cite{Basso2022a, Maziero2023}. We explored in detail the cases of qubits and qutrits. For qubits, the upper bound in terms of the product of initial entanglements is saturated, in agreement with earlier results based on the concurrence measure~\cite{Bergou2021, Maziero2023}. For qutrits, the average distributed entanglement reaches half of the product of the initial entanglements, suggesting that the original bound can be improved. Motivated by these exact results, we proposed a conjecture for an improved upper bound valid for general qudit systems, which becomes exact for low dimensions and for maximally entangled inputs.

Furthermore, we discussed the fact that, although the distributed entanglement may exceed the initial entanglements with some probability---depending on the outcome of the Bell-basis measurement---on average, the entanglement shared between the remote parties cannot surpass the initial entanglement of one of the initial pairs. 

Our analysis highlights how CCRs impose fundamental constraints on the amount of entanglement that can be operationally distributed via ESP. In particular, the local predictability of the subsystems, which quantifies prior information about measurement outcomes, directly limits the entanglement that can be generated through the protocol. This reinforces the idea that entanglement, coherence, and predictability are not independent quantum resources, but are intrinsically linked by complementarity relations.

\vspace*{0.5cm}
\begin{acknowledgments}
\vspace*{-0.4cm}
This work was supported by the Coordination for the Improvement of Higher Education Personnel (CAPES), Grant No. 88887.827989/2023-00, the S\~ao Paulo Research Foundation (FAPESP), Grant No.~2022/09496-8.3, the National Council for Scientific and Technological Development (CNPq), Grants No. 309862/2021-3, No. 409673/2022-6, and No. 421792/2022-1, and the National Institute for the Science and Technology of Quantum Information (INCT-IQ), Grant No. 465469/2014-0. JM gratefully acknowledges the hospitality of Rafael Chaves and the International Institute of Physics in Natal, RN, Brazil, where part of this work was carried out. LCC also acknowledges CNPq, Grant No.~308065/2022-0.
\end{acknowledgments}

%
%


\begin{thebibliography}{9}


\bibitem{Heisenberg1925} W. Heisenberg,
\"Uber quantentheoretische Umdeutung kinematischer und mechanischer Beziehungen., 
Z. Physik \textbf{33}(1), 879 (1925).


\bibitem{Born1925} M. Born and P. Jordan,
Zur Quantenmechanik, 
Z. Physik \textbf{34}(1), 858 (1925).


\bibitem{Dirac1925} P. A. M. Dirac,
The fundamental equations of quantum mechanics,
Proc. R. Soc. A \textbf{109}(752), 642 (1925).


\bibitem{Pauli1925} W. Pauli,
\"Uber den Zusammenhang des Abschlusses der Elektronengruppen im Atom mit der Komplexstruktur der Spektren, 
Zeitschrift f\"{u}r Physik, \textbf{31}(1), 765 (1925).


\bibitem{BornH1926} M. Born, W. Heisenberg, and P. Jordan,
Zur Quantenmechanik. II., 
Z. Physik \textbf{35}(8), 557 (1926).


\bibitem{Schrodinger1926} E. Schr\"{o}dinger,
Quantisierung als Eigenwertproblem,
Annalen der Physik, \textbf{79}, 361 (1926).


\bibitem{Born1926} M. Born,
Quantenmechanik der Sto\ss vorg\"ange,
Zeitschrift f\"{u}r Physik, \textbf{38}(11), 803 (1926).


\bibitem{Born1927} M. Born,
Das Adiabatenprinzip in der Quantenmechanik,
Z. Physik \textbf{40}(3), 167 (1927).


\bibitem{Planck1900} M. Planck,
Zur Theorie des Gesetzes der Energieverteilung im Normalspectrum, 
Verhandlungen der Deutschen Physikalischen Gesellschaft, \textbf{2}, 237 (1900).


\bibitem{Einstein1905} A. Einstein,
\"Uber einen die Erzeugung und Verwandlung des Lichtes betreffenden heuristischen Gesichtspunkt [AdP 17, 132 (1905)],
Annalen Der Physik \textbf{517}, 164 (2005).


\bibitem{Bohr1913} N. Bohr,
I. On the constitution of atoms and molecules, 
Phil. Mag. J. Sci. \textbf{26}(151), 1 (1913).


\bibitem{Broglie1924} L. de Broglie, \textit{Recherches sur la th\'{e}orie des quanta} (Doctoral dissertation, Migration-universit\'{e} en cours d'affectation, 1924).




%
%



\bibitem{Bohr1928} N. Bohr,
The Quantum Postulate and the Recent Development of Atomic Theory,
Nature \textbf{121}(3050), 580 (1928).


\bibitem{Einstein1935} A. Einstein, B. Podolsky, and N. Rosen,
Can Quantum-Mechanical Description of Physical Reality Be Considered Complete?,
Phys. Rev. \textbf{47}(10), 777 (1935).


\bibitem{Schrodinger1935} E. Schr\"{o}dinger,
Die gegenwärtige Situation in der Quantenmechanik. 
Naturwissenschaften, \textbf{23}(50), 844 (1935).


\bibitem{Bell1964} J. S. Bell,
On the Einstein Podolsky Rosen paradox,
Physics Physique Fizika \textbf{1}(3), 195 (1964).


\bibitem{Feynman1982} R. P. Feynman,
Simulating physics with computers, 
Int. J. Theor. Phys. \textbf{21}(6), 467 (1982).


\bibitem{Lloyd1996} S. Lloyd,
Universal Quantum Simulators | 
Science, \textbf{273}(5278), 1073 (1996).


\bibitem{Simon2017} C. Simon,
Towards a global quantum network,
Nature Photon. \textbf{11}(11), 678 (2017).


\bibitem{Ramya2025} R. Ramya, P. Kumar, D. Dhanasekaran, R. S. Kumar, and S. A. Sharavan,
A review of quantum communication and information networks with advanced cryptographic applications using machine learning, deep learning techniques, 
Franklin Open \textbf{10}, 100223 (2025).


\bibitem{Hu2023} X.-M. Hu, Y. Guo, B.-H. Liu, C.-F. Li, and G.-C. Guo, Progress in quantum teleportation,
Nat. Rev. Phys. \textbf{5}(6), 339 (2023).


\bibitem{Bennett1993} C. H. Bennett, G. Brassard, C. Cr\'{e}peau, R. Jozsa, A. Peres, and W. K. Wootters,
Teleporting an unknown quantum state via dual classical and Einstein-Podolsky-Rosen channels, 
Phys. Rev. Lett. \textbf{70}(13), 1895 (1993).


%
%
%


\bibitem{Zukowski1993} M. \.{Z}ukowski, A. Zeilinger, M. A. Horne, and A. K. Ekert,
Event-ready-detectors'' Bell experiment via entanglement swapping,
Phys. Rev. Lett. \textbf{71}(26), 4287 (1993).


\bibitem{Wei2022} S. Wei \textit{et al.},
Towards Real-World Quantum Networks: A Review,
Laser \& Photonics Reviews \textbf{16}(3), 2100219 (2022).


\bibitem{Drmota2023} P. Drmota \textit{et al.},
Robust Quantum Memory in a Trapped-Ion Quantum Network Node,
Phys. Rev. Lett. \textbf{130}(9), 090803 (2023).


\bibitem{Kimble2008} H. J. Kimble,
The quantum internet,
Nature \textbf{453}(7198), 1023 (2008).


\bibitem{Kumar2025} V. Kumar, C. Cicconetti, M. Conti, and A. Passarella,
Quantum Internet: Technologies, Protocols, and Research Challenges, 
arXiv:2502.01653 (2025).


\bibitem{Rohde2025} P. P. Rohde \textit{et al.},
The Quantum Internet (Technical Version), 
arXiv:2501.12107 (2025).


\bibitem{Pan1998} J.-W. Pan, D. Bouwmeester, H. Weinfurter, and A. Zeilinger,
Experimental Entanglement Swapping: Entangling Photons That Never Interacted,
Phys. Rev. Lett. \textbf{80}(18), 3891 (1998)


\bibitem{Jennewein2001} T. Jennewein, G. Weihs, J.-W. Pan, and A. Zeilinger,
Experimental Nonlocality Proof of Quantum Teleportation and Entanglement Swapping,
Phys. Rev. Lett. \textbf{88}(1), 017903 (2001).


\bibitem{Sciarrino2002} F. Sciarrino, E. Lombardi, G. Milani, and F. De Martini,
Delayed-choice entanglement swapping with vacuum--one-photon quantum states,
Phys. Rev. A \textbf{66}(2), 024309 (2002).


\bibitem{Riedmatten2005} H. de Riedmatten, I. Marcikic, J. A. W. van Houwelingen, W. Tittel, H. Zbinden, and N. Gisin,
Long-distance entanglement swapping with photons from separated sources,
Phys. Rev. A \textbf{71}(5), 050302 (2005).


\bibitem{Bose1998} S. Bose, V. Vedral, and P. L. Knight,
Multiparticle generalization of entanglement swapping,
Phys. Rev. A \textbf{57}(2), 822 (1998).


\bibitem{Bose1999} S. Bose, V. Vedral, and P. L. Knight,
Purification via entanglement swapping and conserved entanglement,
Phys. Rev. A \textbf{60}(1), 194 (1999).


\bibitem{Schmid2009} C. Schmid, N. Kiesel, U. K. Weber, R. Ursin, A. Zeilinger, and H. Weinfurter,
Quantum teleportation and entanglement swapping with linear optics logic gates,
New J. Phys. \textbf{11}(3), 033008 (2009).


\bibitem{Sangouard2011} N. Sangouard, C. Simon, H. De Riedmatten, and N. Gisin,
Quantum repeaters based on atomic ensembles and linear optics,
Rev. Mod. Phys. \textbf{83}(1), 33 (2011).


\bibitem{Branciard2012} C. Branciard, D. Rosset, N. Gisin, and S. Pironio,
Bilocal versus nonbilocal correlations in entanglement-swapping experiments,
Phys. Rev. A \textbf{85}(3), 032119 (2012).


\bibitem{Jin2015} R.-B. Jin, M. Takeoka, U. Takagi, R. Shimizu, and M. Sasaki,
Highly efficient entanglement swapping and teleportation at telecom wavelength,
Sci. Rep. \textbf{5}(1), 9333 (2015).


\bibitem{Khalique2015} A. Khalique and B. C. Sanders,
Practical long-distance quantum key distribution through concatenated entanglement swapping with parametric down-conversion sources,
J. Opt. Soc. Am. B \textbf{32}(11) 2382 (2015).


\bibitem{Tsujimoto2018} Y. Tsujimoto \textit{et al.},
High-fidelity entanglement swapping and generation of three-qubit GHZ state using asynchronous telecom photon pair sources,
Sci. Rep. \textbf{8}(1), 1446 (2018).


\bibitem{Basso2022a} M. L. W. Basso and J. Maziero,
Operational connection between predictability and entanglement in entanglement swapping from partially entangled pure states,
Phys. Lett. A, vol. \textbf{451}, 128414 (2022).
%

\bibitem{Maziero2023} J. Maziero, M. L. W. Basso, and L. C. C\'{e}leri,
Local predictability and coherence versus distributed entanglement in entanglement swapping from partially entangled pure states,
Phys. Lett. A \textbf{457}, 128576 (2023).


\bibitem{Davis2025} S. I. Davis \textit{et al.},
Entanglement Swapping Systems toward a Quantum Internet,
arXiv:2503.18906 (2025).


\bibitem{Bouda2001} J. Bouda and V. Buzek,
Entanglement swapping between multi-qudit systems,
J. Phys. A \textbf{34}(20), 4301 (2001).


\bibitem{Bergou2021} J. A. Bergou, D. Fields, M. Hillery, S. Santra, and V. S.
Malinovsky,
Average concurrence and entanglement swapping,
Phys. Rev. A \textbf{104}(2), 022425 (2021).


\bibitem{Hurst1984} Hurst,
Multiple-Valued Logic - its Status and its Future,
IEEE Transactions on computers, \textbf{100}(12), 1160 (1984).


\bibitem{Bormashenko2019} E. Bormashenko,
Generalization of the Landauer Principle for Computing Devices Based on Many-Valued Logic,
Entropy \textbf{21}(12), 1150 (2019).


\bibitem{Sandhie2021} Z. T. Sandhie, J. A. Patel, F. U. Ahmed, and M. H. Chowdhury,
Investigation of multiple-valued logic technologies for beyond-binary era,
ACM Comp. Sur. \textbf{54}(1), 1 (2021).

%
\bibitem{Zhu2025} X. Zhu \textit{et al.},
High-performance ternary logic circuits and neural networks based on carbon nanotube source-gating transistors,
Sci. Adv. \textbf{11}(2), eadt1909 (2025).



\bibitem{Farias2025} T. de S. Farias, L. Friedrich, and J. Maziero,
A Short Review on Qudit Quantum Machine Learning,
arXiv:2505.05158 (2025).


\bibitem{Balantekin2024} A. B. Balantekin and A. M. Suliga,
On the properties of qudits,
Eur. Phys. J. A \textbf{60}(6), 124 (2024).


\bibitem{Di2015} Y.-M. Di and H.-R. Wei,
Optimal synthesis of multivalued quantum circuits,
Phys. Rev. A \textbf{92}(6), 062317 (2015).


\bibitem{Wang2020} Y. Wang, Z. Hu, B. C. Sanders, and S. Kais,
Qudits and High-Dimensional Quantum Computing,
Front. Phys. \textbf{8}, 589504 (2020).


\bibitem{Nikolaeva2024} A. S. Nikolaeva, E. O. Kiktenko, and A. K. Fedorov,
Efficient realization of quantum algorithms with qudits,
EPJ Quantum Technol. \textbf{11}(1), 1 (2024).


\bibitem{Lanyon2009} B. P. Lanyon \textit{et al.},
Simplifying quantum logic using higher-dimensional Hilbert spaces,
Nature Phys. \textbf{5}(2), 134 (2009).


\bibitem{Duclos-Cianci2013} G. Duclos-Cianci and D. Poulin,
Kitaev's ${\mathbb{Z}}_{d}$-code threshold estimates,
Phys. Rev. A \textbf{87}(6), 062338 (2013).


\bibitem{Watson2015} F. H. E. Watson, H. Anwar, and D. E. Browne,
Fast fault-tolerant decoder for qubit and qudit surface codes,
Phys. Rev. A \textbf{92}(3), 032309 (2015).


\bibitem{Campbell2012} E. T. Campbell, H. Anwar, and D. E. Browne,
Magic-State Distillation in All Prime Dimensions Using Quantum Reed-Muller Codes,
Phys. Rev. X \textbf{2}(4), 041021 (2012).


\bibitem{Anwar2014} H. Anwar, B. J. Brown, E. T. Campbell, and D. E. Browne,
Fast decoders for qudit topological codes,
New J. Phys. \textbf{16}(6), 063038 (2014).


\bibitem{Anwar2012} H. Anwar, E. T. Campbell, and D. E. Browne,
Qutrit magic state distillation,
New J. Phys. \textbf{14}(6), 063006 (2012).


\bibitem{Campbell2014} E. T. Campbell,
Enhanced Fault-Tolerant Quantum Computing in $d$-Level Systems,
Phys. Rev. Lett. \textbf{113}(23), 230501 (2014).


\bibitem{Andrist2015} R. S. Andrist, J. R. Wootton, and H. G. Katzgraber,
Error thresholds for Abelian quantum double models: Increasing the bit-flip stability of topological quantum memory,
Phys. Rev. A \textbf{91}(4), 042331 (2015).


\bibitem{Chizzini2022} M. Chizzini \textit{et al.},
Quantum error correction with molecular spin qudits,
Phys. Chem. Chem. Phys. \textbf{24}(34), 20030 (2022).


\bibitem{Fischer2022} L. E. Fischer, D. Miller, F. Tacchino, P. Kl. Barkoutsos, D. J. Egger, and I. Tavernelli,
Ancilla-free implementation of generalized measurements for qubits embedded in a qudit space,
Phys. Rev. Res. \textbf{4}(3), 033027 (2022).


\bibitem{Stricker2022} R. Stricker \textit{et al.},
Experimental Single-Setting Quantum State Tomography,
PRX Quan. \textbf{3}(4), 040310 (2022).


\bibitem{Weggemans2022} J. R. Weggemans \textit{et al.},
Solving correlation clustering with QAOA and a Rydberg qudit system: a full-stack approach,
Quantum \textbf{6}, 687 (2022).


\bibitem{Ringbauer2022} M. Ringbauer \textit{et al.},
A universal qudit quantum processor with trapped ions,
Nat. Phys. \textbf{18}(9), 1053 (2022).


\bibitem{Hrmo2023} P. Hrmo \textit{et al.},
Native qudit entanglement in a trapped ion quantum processor,
Nat. Commun. \textbf{14}(1), 2242 (2023).


\bibitem{Chi2022} Y. Chi \textit{et al.},
A programmable qudit-based quantum processor,
Nat. Commun. \textbf{13}(1), 1166 (2022).


\bibitem{Lu2020} H.-H. Lu \textit{et al.},
Quantum Phase Estimation with Time-Frequency Qudits in a Single Photon,
Adv. Quan. Tech. \textbf{3}(2), 1900074 (2020).


\bibitem{Gonzalez-Cuadra2022} D. Gonz\'{a}lez-Cuadra, T. V. Zache, J. Carrasco, B. Kraus, and P. Zoller,
Hardware Efficient Quantum Simulation of Non-Abelian Gauge Theories with Qudits on Rydberg Platforms,
Phys. Rev. Lett. \textbf{129}(16), 160501 (2022).


\bibitem{Cohen2021} S. R. Cohen and J. D. Thompson,
Quantum Computing with Circular Rydberg Atoms,
PRX Quan. \textbf{2}(3), 030322 (2021).


\bibitem{Ahn2000} J. Ahn, T. C. Weinacht, and P. H. Bucksbaum,
Information storage and retrieval through quantum phase,
Science \textbf{287}(5452), 463 (2000),


\bibitem{Fischer2023} L. E. Fischer, A. Chiesa, F. Tacchino, D. J. Egger, S. Carretta, and I. Tavernelli,
Universal Qudit Gate Synthesis for Transmons,
PRX Quan. \textbf{4}(3), 030327 (2023),


\bibitem{Morvan2021} A. Morvan \textit{et al.},
Qutrit Randomized Benchmarking,
Phys. Rev. Lett. \textbf{126}(21), 210504 (2021),


\bibitem{Kasper2021} V. Kasper \textit{et al.},
Universal quantum computation and quantum error correction with ultracold atomic mixtures,
Quan. Sci. Technol. \textbf{7}(1), 015008 (2021),


\bibitem{Smith2013} A. Smith, B. E. Anderson, H. Sosa-Martinez, C. A. Riofr\'{i}o, I. H. Deutsch, and P. S. Jessen,
Quantum Control in the Cs $6{S}_{1/2}$ Ground Manifold Using Radio-Frequency and Microwave Magnetic Fields,
Phys. Rev. Lett. \textbf{111}(17), 170502 (2013),


\bibitem{Anderson2015} B. E. Anderson, H. Sosa-Martinez, C. A. Riofr\'{i}o, I. H. Deutsch, and P. S. Jessen,
Accurate and Robust Unitary Transformations of a High-Dimensional Quantum System,
Phys. Rev. Lett. \textbf{114}(24), 240401 (2015),


\bibitem{Senko2015} C. Senko \textit{et al.},
Realization of a Quantum Integer-Spin Chain with Controllable Interactions,
Phys. Rev. X \textbf{5}(2), 021026 (2015).


\bibitem{Deller2023} Y. Deller \textit{et al.},
Quantum approximate optimization algorithm for qudit systems,
Phys. Rev. A \textbf{107}(6), 062410 (2023).


\bibitem{MacDonell2021} R. J. MacDonell \textit{et al.},
Analog quantum simulation of chemical dynamics,
Chem. Sci. \textbf{12}(28), 9794 (2021).


\bibitem{Cao2019} Y. Cao \textit{et al.},
Quantum Chemistry in the Age of Quantum Computing,
Chem. Rev. \textbf{119}(19), 10856 (2019).


\bibitem{Maskara2025} N. Maskara \textit{et al.},
Programmable simulations of molecules and materials with reconfigurable quantum processors,
Nat. Phys. \textbf{21}(2), 289 (2025).


\bibitem{McArdle2020} S. McArdle, S. Endo, A. Aspuru-Guzik, S. C. Benjamin, and X. Yuan,
Quantum computational chemistry,
Rev. Mod. Phys. \textbf{92}(1), 015003 (2020).


\bibitem{Haldane1983} F. D. M. Haldane,
Nonlinear Field Theory of Large-Spin Heisenberg Antiferromagnets: Semiclassically Quantized Solitons of the One-Dimensional Easy-Axis N\'eel State,
Phys. Rev. Lett. \textbf{50}(15), 1153 (1983).


\bibitem{Sawaya2020} N. P. D. Sawaya, T. Menke, T. H. Kyaw, S. Johri, A. Aspuru-Guzik, and G. G. Guerreschi,
Resource-efficient digital quantum simulation of $d$-level systems for photonic, vibrational, and spin-s Hamiltonians,
npj Quan. Inf. \textbf{6}(1), 1 (2020).


\bibitem{Wecker2015} D. Wecker, M. B. Hastings, N. Wiebe, B. K. Clark, C. Nayak, and M. Troyer,
Solving strongly correlated electron models on a quantum computer,
Phys. Rev. A \textbf{92}(6), 062318 (2015).


\bibitem{Rico2018} E. Rico \textit{et al.},
SO(3) `Nuclear Physics' with ultracold Gases,
Annals of Phys. \textbf{393}, 466 (2018).


\bibitem{Meth2025} M. Meth \textit{et al.},
Simulating two-dimensional lattice gauge theories on a qudit quantum computer,
Nat. Phys. \textbf{21}(4), 570 (2025).


\bibitem{Basso2020a} M. L. W. Basso and J. Maziero,
Complete complementarity relations for multipartite pure states,
J. Phys. A \textbf{53}(46), 465301 (2020).


\bibitem{Basso2021} M. L. W. Basso and J. Maziero,
Complete complementarity relations and their Lorentz invariance, 
Proc. R. Soc. A. {\bf 477}(2253), 20210058 (2021).


\bibitem{Basso2022b} M. L. W. Basso and J. Maziero,
Entanglement Monotones from Complementarity Relations,
J. Phys. A {\bf 55}(35), 355304 (2022).


\bibitem{Jakob2010} M. Jakob and J. A. Bergou,
Quantitative complementarity relations in bipartite systems: Entanglement as a physical reality,
Opt. Commun. \textbf{283}(5), 827 (2010).


\bibitem{Araujo2025} J. S. Ara\'ujo, D. S. Starke, A. S. Coelho, J. Maziero, G. H. Aguilar, and R. M. Angelo,
Quantum observable's reality erasure with spacelike-separated operations, 
Phys. Rev. A \textbf{111}(6), 062425 (2025).


\bibitem{Basso2020b} M. L. W. Basso, D. S. S. Chrysosthemos, and J. Maziero,
Quantitative Wave-Particle Duality Relations from the Density Matrix Properties,
Quan. Inf. Process. {\bf 19}(8), 254 (2020).


\bibitem{Starke2024} D. S. Starke, M. L. W. Basso, and J. Maziero,
An updated quantum complementarity principle,
Proc. R. Soc. A \textbf{480}(2301), 20240517 (2024).




\bibitem{bennett93} C. H. Bennett, G. Brassard, C. Cr\'epeau, R. Jozsa, A. Peres, and W. K. Wootters,
Teleporting an unknown quantum state via dual classical and Einstein-Podolsky-Rosen channels
Phys. Rev. Lett. {\bf 70}(13), 1895~(1993).


\bibitem{Karimipour2002} V. Karimipour, A. Bahraminasab, and S. Bagherinezhad,
Quantum key distribution for $d$-level systems with generalized Bell states,
Phys. Rev. A \textbf{65}(5), 052331 (2002).


\bibitem{Sych2009} D. Sych and G. Leuchs,
A complete basis of generalized Bell states,
New J. Phys. \textbf{11}(1), 013006 (2009).


\bibitem{Camps2021} D. Camps, R. Van Beeumen, and C. Yang,
Quantum Fourier transform revisited,
Numerical Linear Algebra with Applications \textbf{28}(1), e2331 (2021).




\end{thebibliography}
\end{document}